\documentclass[11pt]{article}               
 
\usepackage{amssymb}  
\usepackage{amscd} 
\usepackage{xypic} 
\usepackage[matrix,arrow,curve]{xy}             
 
\newcommand {\newsection}{\setcounter{equation}{0}\section}

\def \rr {{\mathbb R}} 
\def \de {\partial} 
 
\def \Ka {K\" ahler}

\newcommand{\al}{\alpha} 
 
\newcommand{\ga}{\gamma} 
\newcommand{\eps}{\epsilon} 
\newcommand{\om}{\omega}

\newcommand{\G}{\Gamma}

\newcommand{\beq}{\begin{equation}} 
\newcommand{\eeq}{\end{equation}} 
\newcommand{\bea}{\begin{eqnarray}} 
\newcommand{\eea}{\end{eqnarray}} 
\newcommand{\bean}{\begin{eqnarray*}} 
\newcommand{\eean}{\end{eqnarray*}} 
\newcommand{\non}{\nonumber} 
\newcommand{\baa}{\bar{a}} 
\newcommand{\bbb}{\bar{b}} 
\newcommand{\bc}{\bar{c}}

\newcommand{\bz}{\bar{z}}

\newcommand{\tss}{\tilde{s}}

\newcommand{\he}{\hat{e}}

\newcommand{\hD}{\hat{D}}

\newcommand{\hGa}{\hat{\Gamma}} 
\newcommand{\haom}{\hat{\omega}}

\begin{document} 
\setcounter{page}{0} 
\begin{titlepage}    
\titlepage 
\rightline{hep-th/0206213} 
\rightline{CPHT-RR 039.0502} 
\vskip 3cm 
\centerline{{ \bf \large Kaluza-Klein bundles and manifolds of 
exceptional holonomy}} 
\vskip 1.5cm 
\centerline{Peter Kaste, Ruben Minasian, Michela Petrini and  Alessandro 
Tomasiello} 
\begin{center} 
\em Centre de Physique Th{\'e}orique, Ecole 
Polytechnique\footnote{Unit{\'e} mixte du CNRS et de l'EP, UMR 
7644} 
\\91128 Palaiseau Cedex, France\\{\tt 
\vskip .5cm 
kaste, ruben, petrini, tomasiel@cpht.polytechnique.fr} 
\end{center} 
\vskip 1.5cm 
\begin{abstract} 
 
We show how in the presence of RR two-form field strength 
the conditions for preserving supersymmetry 
on six- and seven-dimensional manifolds lead to certain 
generalizations of monopole equations.  
For six dimensions the string frame metric is K\"ahler with the 
complex structure that descends from the octonions
if in addition we assume $F^{(1,1)}=0$. The susy generator 
is a gauge covariantly constant spinor. For seven dimensions the 
string frame metric is conformal to a $G_2$ metric if 
in addition we assume
the field strength to obey a selfduality constraint. 
Solutions to these equations 
lift to geometries of $G_2$ and $Spin(7)$ holonomy respectively.

\end{abstract} 
 
\vfill 
\begin{flushleft} 
{\today}\\ 
%\vspace{.5cm} 
\end{flushleft} 
\end{titlepage}  
 
\newpage 
 
\newsection{Introduction} 
 
Recent progress in understanding ${\cal N}=1$ dynamics  
has to a large extent relied on exploiting dual realizations of such  
theories. A prominent example of such dual pairs are 
D6 branes of type IIA string theory wrapping supersymmetric three cycles  
in a Calabi-Yau threefold on one side  
and M-theory compactifications on manifolds of 
$G_2$ holonomy on the other side, \cite{Atiyah:2000zz,Acharya:2000gb}. 
A less studied example realizing such a duality for ${\cal N}=1$  
theories in three dimensions are   
D6 branes wrapping supersymmetric four cycles  
in a $G_2$ manifold and M-theory compactifications on manifolds of 
$Spin(7)$ holonomy.  
Together with D6-branes in flat space, this exhausts the list of 
spacetime filling D6-branes partially wrapped on supersymmetric cycles 
of an internal $d$-dimensional space $X$ that lift in M-theory to 
compactifications on $(d+1)$-dimensional spaces $Y$ preserving half 
as many supersymmetries as does $X$,  
\beq 
\begin{array}{|c||c|c|c|} \hline  
{} & {\rm Hol}(X) & \mbox{susy cycle} & {\rm Hol}(Y) \\ \hline\hline 
d=3 & \{ 1 \} & \{ pt \} & SU(2) \\ 
d=6 & SU(3)  & {\rm SLAG} & G_2 \\ 
d=7 & G_2   & {\rm coassociative} & Spin(7) \\ \hline 
\end{array}\ . \label{list} 
\eeq 
In all these cases, the power of the M-theory realizations resides in the 
fact that they are completely geometrical, no background fluxes being 
switched on.  
Much progress has been achieved recently in the study of these M-theory 
constructions as well as in the derivation of explicit metrics for certain 
non-compact manifolds of exceptional holonomy (for a review with an 
extensive list of references see e.g. \cite{Duff}).  
 
Our purpose here is a 
further clarification of the connections between the geometries and associated 
structures in $d$ and $d+1$ dimensions, as well as the general properties 
of the Kaluza-Klein (monopole)  bundles. The strategy is rather 
conventional -- we will analyze the conditions for preserving supersymmetry 
in type IIA string theory with non\-trivial RR vector field and  
dilaton, namely all the fields coming from the eleven-dimensional metric. 
The result is a set of gauge equations in six and seven dimensions 
which are close relatives to the familiar three-dimensional monopole 
equations that arise in the flat case; schematically they take 
the form  
\beq 
\de_a \phi \sim \eta_{abc}F^{bc} , 
\eeq 
where 
in six dimensions the indices are holomorphic and the 
form $\eta$ is only of type $(3,0)$, whereas in seven dimensions the 
indices span all of the seven dimensions and the three-form is the 
$G_2$ form $\Phi$.  
 
For D6-branes on special Lagrangian three-cycles in a non-compact 
Calabi-Yau threefold, we also show that the almost complex structure 
descending from the octonions is integrable and that the ten-dimensional 
string frame metric in the presence of the D6-branes together with 
this complex structure makes the internal space into a K\"ahler 
manifold, if the further constraint $F^{(1,1)}=0$ is imposed, which
is stronger than the condition $F^{ab}J_{ab}=0$, required by
supersymmetry. 
The ten-dimensional string frame metric is a warped product 
whose internal part is this K\"ahler metric. 
The supersymmetry generator, however, is then a {\em gauge 
  covariantly constant} spinor.  In seven dimensions, by imposing the  
monopole equations
we find that, in the presence of the D6-branes,
 the string frame metric on the internal manifold is conformal 
to a metric of $G_2$ holonomy, if in addition a selfduality of the 
field strength is assumed. 
 
The structure of the paper is as follows.  
In section \ref{sec.flat} we review the case of D6-branes in flat space.  
In section \ref{sec.g2su3} we then turn to supersymmetric D6-branes on 
Calabi-Yau manifolds where we present three alternative derivations of the 
monopole equations and the geometry in the string frame metric. First 
we use the supersymmetry constraints in type IIA in the presence of 
nontrivial dilaton and RR two-form field strength. Then we derive
the same result from the existence of a $G_2$ structure on the lift 
(see \cite{Cvetic:2002da} for a similar ansatz) and lastly from the 
existence of a selfdual spin connection on the lift 
\cite{Bilal:2001an}.  In section 
\ref{sec.spin7g2} the same line of thought as in section  
\ref{sec.g2su3} is applied to  
a background of D6-branes on supersymmetric four-cycles in  
$G_2$ manifolds that lift to $Spin(7)$ manifolds. In section \ref{sec.disc} 
we  summarize and discuss our results. 
 
\section{Review of D6-branes in flat space, or: $SU(2)\to\{ 1 \}$} 
\label{sec.flat} 
 
Before we start to analyze supersymmetric D6-branes in Calabi-Yau manifolds,  
let us  recall the well-understood case of $N$ coincident 
D6-branes of type IIA in flat ten-dimensional space.  These branes are 
magnetically charged under the RR one-form potential $A$. 
This potential can be identified with the connection of the 
principal U(1)-bundle over $\rr^3-\{0\}$, the transverse space to the  
D6-branes, describing a magnetic monopole of 
charge $N$. The associated Maxwell equations imply that  
\beq 
\de_i V(x) = -\frac{1}{2} \eps_{ijk} F^{jk}(x) \quad 
\Leftrightarrow \quad  
dV = - \tilde{*} F, \qquad \mbox{where}~~ 
V(x)= \epsilon + {N \over 4 \pi |\, x\, |}   
\label{mono} 
\eeq 
with $F=dA$, an arbitrary 
integration constant $\epsilon$ and where the Hodge * is taken w.r.t.\  
the flat metric on $\rr^3$.  
The scalar potential $V(x)$ is  
harmonic on $\rr^3-\{0\}$ with  
$\Delta V(x)=-\frac{1}{2} \eps_{ijk} \de ^iF^{jk}(x)=-N \delta(x)$, 
where the $\delta$-function indicates the presence of the monopoles. 
Solutions for the potential $A$ are given by the well-known Wu-Yang 
monopole potentials. 
 
When lifted to eleven dimensions, the configuration becomes purely  
geometrical and the magnetic monopole in three dimensions a gravitational  
instanton in four dimensions. The  
additional M-theory circle is loosely speaking the fiber of the monopole 
bundle. To be more precise, the four-dimensional transverse space has a 
metric \cite{Hawking:1976jb,Eguchi:1978gw} 
\beq 
ds^2_{4} = V(x) d\tss_3^2 + V^{-1}(x) (dz + A)^2,  
\label{ds4} 
\eeq 
where $z$ is a periodic coordinate and $d\tss_3^2$ the Euclidean metric 
on $\rr^3$.   
The monopole equation (\ref{mono}) is precisely the requirement of  
anti-selfdual spin connection of the metric (\ref{ds4}), which implies 
Ricci-flatness. 
The metric is an example of Hawking's 
multi-center metrics for gravitational  
instantons in a limit where the $N$ centers coincide.  
The non-negative integration 
constant $\epsilon$ controls the asymptotic behavior of the circle 
parameterized by $z$. For $\eps=0$ it decompactifies and the metric 
becomes asymptotically locally Euclidean. For $\eps=0$ and $N=2$ it is 
the Eguchi-Hanson metric. For nonvanishing $\eps$ the circle remains 
compact also asymptotically. For $\eps=1$ and $N=1$ it is the  
Taub-NUT metric. 
 
The background metric for the M-theory compactification is thus 
\beq 
ds_{11}^2 =  e ^{-2\al \phi} ds_{10}^2 + e ^{2\beta \phi}(dz+A)^2 
 = d\tss_7^2 + ds_4^2,   
\label{ds11.flat} 
\eeq   
where $d\tss_7^2$ is a flat Minkowski metric in seven dimensions,  
$ds_4^2$ is given in (\ref{ds4}),  
$V=e^{-2\beta\phi}$ and where $ds_{10}^2$ is 
the physical metric in ten dimensions.   
The parameters $\al$ and $\beta$ determine the frame in which this 
metric is given. For a string frame in ten dimensions they are  
$(\al,\beta)=(1/3,2/3)$, but for comparability with different frames 
we will leave them general in most of the equations unless we refer 
explicitly to the ten-dimensional string frame.  
The physical ten-dimensional metric is thus a warped product 
\beq  
ds_{10}^2 = e ^{2\al \phi} d\tss_7^2 + e ^{2(\al-\beta) \phi} d\tss_{3}^2   
\label{ds10.flat} 
\eeq   
of two flat metrics. I.e. the presence of the D6-branes shows up in the  
string frame metric as a warping by $V^{-1/2}$ and $V^{1/2}$ of the 
longitudinal and transverse directions, respectively. Using the string frame  
relation $\beta=2\al$, the monopole equation (\ref{mono}) can be rewritten 
as 
\beq 
dV=d\left( e^{-2\beta\phi} \right) = -*\left( e^{-\al\phi}F \right), 
\label{mono.flat.string} 
\eeq 
where the Hodge * is now taken w.r.t. the string frame metric  
$e ^{-2\al \phi} d\tss_{3}^2$ on the internal space.

\newsection{$G_2\to SU(3)$} 
\label{sec.g2su3} 
 
In this section we will analyze the geometry of 
D6-branes wrapped on supersymmetric cycles in  
non-compact Calabi-Yau manifolds. 
As mentioned in the introduction, we start from the examination of the 
conditions for preserving supersymmetry in type IIA string theory in 
the presence of the RR vector field and a nontrivial dilaton. 
It is easiest to derive these conditions from  
reduction of the supersymmetry conditions of eleven-dimensional 
supergravity. To this end our 
conventions are as follows. The eleven-dimensional metric 
\beq 
ds_{11}^2= e ^{-2\al \phi} ds_{10}^2 + e ^{2\beta \phi}(dz+A)^2 
 = d\tss_4^2 + ds_7^2   
\label{ds11} 
\eeq  
is assumed to be the direct product of a Minkowski metric 
$d\tss_4^2$ with a non-compact $G_2$ metric $ds_7^2$, the latter having 
a $U(1)$ 
isometry parameterized by the coordinate $z$.  
We assume that the field strength of the KK-vector $A$ has nonvanishing 
components only in the internal dimensions. Also the dilaton $\phi$  
depends nontrivially only on the internal coordinates. 
The metric $ds_{10}^2$ is  
the physical metric in ten dimensions;    
the parameters $\al$ and $\beta$ determine the frame, with 
$(\al,\beta)=(1/3,2/3)$ for the string frame in ten dimensions.  
The physical ten-dimensional metric is thus of a warped type 
\beq  
ds_{10}^2 = e ^{2\al \phi} d\tss_4^2 + ds_{6}^2   
\label{ds10} 
\eeq  
and the $G_2$ metric reads 
\beq 
ds_{7}^2 = e ^{-2\al\phi}ds_{6}^2 + e ^{2\beta \phi}(dz+A)^2. 
\label{ds7} 
\eeq 
 
We will denote objects referring to the metric $ds_7^2$ with hats, 
whereas the others refer to the metric $ds_6^2$. 
Upper case frame indices run over the range $A,B,C=1,\ldots,7$,  
where the index 7 refers to the $z$-direction,  
lower case frame indices have the range $a,b,c=1,\ldots,6$.  
 
\subsection{Supersymmetry and holomorphic monopoles} 
 
With our assumptions on the eleven-dimensional metric (\ref{ds11}) and 
the field strength being internal, the condition for ${\cal N}=1$ 
supersymmetry in four dimensions reduces to the condition of having 
exactly one covariantly constant  
Majorana spinor on the internal seven-manifold 
with metric (\ref{ds7}), 
\beq 
\hD_A \eps = \left( \hat{\partial}_A +\frac{1}{4} \hGa_{BAC} \ga ^{BC} 
\right) \eps =0.  
\label{cov7} 
\eeq 
Using the following relations between the spin connection coefficients 
corresponding to the metrics $ds_7^2$ and $ds_6^2$,   
\beq 
\begin{array}{l} 
\vspace{.2cm}\hat\Gamma_{abc} = e^{\al\phi} \left\{ \G_{abc} + 
 \al \left[ \delta_{bc} (\de_a \phi)- \delta_{ba} (\de_c \phi) 
 \right] \right\} \ , \\\vspace{.2cm} 
\hat\Gamma_{a\, z \, c} = - \frac{1}{2} e^{(2\al+\beta)\phi} F_{ac} \ , 
\\ \vspace{.2cm}  
\hat\Gamma_{z \, bc}  =  - \frac{1}{2} e^{(2\al+\beta)\phi} F_{bc} \ , 
\\  
\hat\Gamma_{z\, z \, c}  =  \beta e^{\al \phi} (\de_c \phi) \ , 
\end{array} 
\eeq 
together with the constraint that $\eps$ does not depend on $z$, the 
supersymmetry condition (\ref{cov7}) reduces to the  
following system in six dimensions 
\bea 
&&e^{\al\phi} 
\left(D_a + \frac{1}{2}\al (\de_b \phi) \,\ga ^b_{\ a} 
 + \frac14 \check{F} _{ab} 
\ga ^b  \ga \right)\eps = 0, \label{6dgravitino}\\ 
&&-\frac12 e^{\al\phi} 
\left( \frac14 \check{F} _{ab} \ga ^{ab} + \beta \,(\de_a \phi) \,\ga ^a 
\ga \right) 
\eps =0, \label{6ddilatino} 
\eea 
where we have defined  
\[ 
D_a = \de_a +\frac{1}{4} \G_{bac} \ga ^{bc}, \qquad 
\ga = \ga ^7, \qquad\mbox{and}\quad 
\check{F}_{ab} = e ^{(\al+\beta)\phi}F_{ab}. 
\]

We can now turn to possible solutions to (\ref{6dgravitino} - 
\ref{6ddilatino}).  
We are interested in six-dimensional geometries that are related via 
Kaluza-Klein reduction to non-compact $G_2$ manifolds 
with metric (\ref{ds7}).  
The spinor $\eps$ is then the covariantly constant  
Majorana spinor on 
the $G_2$ manifold, satisfying the identity \cite{Marino:2000af} 
\beq  
\ga_{AB}\, \eps = i\,\Phi_{ABC} \,\ga ^C \eps , \label{7dga} 
\eeq 
where $\Phi_{ABC}$ are the structure constants of the imaginary 
octonions. Defining 
$\psi_{abc}\equiv\Phi_{abc}$ and $J_{ab}\equiv\Phi_{ab7}$, this identity 
reduces from the six-dimensional perspective to  
\beq 
\ga_{ab}\eps = i\,\psi_{abc} \ga ^c \eps + i\,J_{ab} \ga \,\eps \ , \qquad  
\ga_a \ga \,\eps  = -i\,J_{ab} \ga ^b \eps ,  
\label{6dga} 
\eeq 
where in addition 
\beq 
J_{a}^{~b}J_{b}^{~c}=-\delta_{a}^{~c}. \label{JJ} 
\eeq 
The last identity implies that $J_a^{~b}$ defines an almost complex structure 
in six dimensions. Due to its antisymmetry, the Riemannian metric is 
actually hermitian w.r.t.\ $J_a^{~b}$.

Plugging (\ref{6dga}) into equations 
(\ref{6dgravitino} - \ref{6ddilatino}), 
we get 
\bea 
&&\left( D_a + \frac{i}{2}\al\, (\de_b \phi)J_{~a}^{b} \,\ga\right) \eps + 
i\, \left( \frac{1}{2}\al\,(\de_b \phi)\, \psi ^b_{\ ac}  
- \frac 14 \check{F}_{ab} J^b_{\ c}\right) \ga ^c \eps = 0 \ , 
\label{6dgrav}\\ 
&&\left( \frac i4 \check{F}_{ab}J^{ab}\right) \ga \eps +   
\left( \frac i4 \check{F}_{ab}\psi^{ab}_{\ \ c}  
-i\, \beta \,(\de_a \phi) J^{a}_{\ c}\right) \ga ^c \eps = 0 \ . 
\label{6ddil} 
\eea 
Here we have dropped the overall factors in front of 
(\ref{6dgravitino} - \ref{6ddilatino}); this means that all the 
equations we get afterwards have to be satisfied outside the locus in 
which $\phi=-\infty$.  
 
Since the spinors $\ga^A\eps$ are linearly independent for $A=1,\ldots,7$, 
\cite{Marino:2000af}, we can conclude that each of the two brackets in  
(\ref{6ddil}) has to vanish, 
\bea 
F^{ab}J_{ab} &=& 0, \label{monopole.1} \\ 
\beta (\de_a \phi)J^a_{~c} &=& \frac{1}{4}\check{F}^{ab}\psi_{abc} 
\qquad \Leftrightarrow \qquad  
d\left(e^{-2\beta\phi}\right)=- *\left(e^{-\al\phi}\psi_3 \wedge F \right).
\label{monopole.2} 
\eea 
Here we have defined the three-form  
$\psi_3=\frac{1}{3!}\psi_{abc}e^a e ^b e ^c$ 
using the orthonormal (w.r.t. $ds_6^2$) cotangent frame $e^a$  
and  the Hodge * is taken w.r.t.\ the string frame metric $ds_6^2$  
on the internal space. The equivalence in (\ref{monopole.2}) uses 
the multiplication properties of the octonionic structure constants (a 
collection of which can e.g.\ be found in the appendix of 
\cite{Bilal:2001an}, whose conventions we follow).  
Using these identities one easily shows that, in the string
frame $\beta=2\al$, (\ref{monopole.2})  
reduces the gravitino equation (\ref{6dgrav}) to
\beq
\left( D_a + \frac{i}{2}\al\, (\de_b \phi)J_{~a}^{b} \,\ga
 -\frac{i}{8} \left[\check{F}_{ab}J^b_{~c}+\check{F}_{cb}J^b_{~a}
 \right] \ga ^c \right) \eps =0 . \label{6dgrav.b}
\eeq 
Splitting the (co)tangent bundle into a holomorphic and an 
anti-holomorphic one w.r.t.\ $J_a^{~b}$, one sees that only the $(1,1)$ 
part of $F$ contributes to the square bracket in (\ref{6dgrav.b}).
Before moving on to analyze this equation, we note that by
taking the dilaton to be constant, in addition to (\ref{monopole.1}) we
get $F^{(2,0)}=0$, thus yielding Hermitian YM equations. The
six-dimensional structures and the role of $F^{(1,1)}$  in such a case are
studied in \cite{CS}.

Moreover it is also the $(1,1)$ part of $F$ that controls the
non-integrability of the almost complex structure. 
Using its representation 
$J_a^{~b}=-i\eps^{\dagger}\ga_a^{~b}\ga \eps$ and the gravitino
equation, it follows that the Nijenhuis tensor takes the form
\beq
N^a_{bc}=\frac{1}{2} \left[ J_c^{~d}\check{F}_{de}\psi ^{e~a}_{~b}
 - J_b^{~d}\check{F}_{de}\psi ^{e~a}_{~c}
 - \check{F}_{bd}(*\psi)^{d~a}_{~c}
 + \check{F}_{cd}(*\psi)^{d~a}_{~b} \right] , \label{Nijenhuis}
\eeq    
where $(*\psi)_{abc}=\frac{1}{3!}\eps_{defabc}\psi^{def}$ are the
components of $*\psi_3$. In the holomorphic/antiholomorphic basis,
labeled by $(a,\bar{a})$, we
can use that $J_a^{\ b}= i\, \delta_a^{\ b}$ and that  
$\psi_{abc}= \frac12 \eps_{abc}$, $\psi_{\bar a\bar b\bar c}=  
\frac12 \eps_{\bar a\bar b\bar c}$, (see (\ref{forms.hol}) and the
remark following it), to find
that the only nonvanishing components of the Nijenhuis tensor are
\beq
N^{\bar{a}}_{bc} = \frac{i}{2} 
 \left( \check{F}_{c\bar{d}}\, \epsilon^{\bar{d}~\bar{a}}_{~b}
       -\check{F}_{b\bar{d}}\, \epsilon^{\bar{d}~\bar{a}}_{~c} \right)
\qquad \mbox{and} \qquad
N^{a}_{\bar{b}\bar{c}} = -\frac{i}{2} 
 \left( \check{F}_{\bar{c}d}\, \epsilon^{d~a}_{~\bar{b}}
       -\check{F}_{\bar{b}d}\, \epsilon^{d~a}_{~\bar{c}} \right)
     . \label{Nijenhuis.hol}
\eeq
The almost complex structure that descends from the
octonions is thus integrable if and only if
\beq 
F^{(1,1)}=0, 
\label{f11} 
\eeq 
which is of course stronger than (\ref{monopole.1}). 
In equation (\ref{CGLP}) we will actually show that the string frame 
metric $ds_6^2$ is then a K\"ahler metric w.r.t.\ to this complex
structure.

In the rest of the paper we will assume the stronger condition
(\ref{f11}) to hold. Together  
with (\ref{monopole.2}) this reduces the system  
(\ref{6dgrav} - \ref{6ddil}) to the condition 
that $\eps$ is a {\em gauge covariantly constant} spinor in six dimensions, 
\beq 
\left( D_a + \frac{i}{2}\al\, (\de_b \phi)J_{~a}^{b} \,\ga\right) \eps  
=0 \quad \Leftrightarrow \quad 
\left\{ \begin{array}{r@{~=~0,}} 
\left( D_a + \frac{i}{2}\al\, (\de_b \phi)J_{~a}^{b} \right) \eps_+ 
\\ 
\left( D_a - \frac{i}{2}\al\, (\de_b \phi)J_{~a}^{b} \right) \eps_- 
\rule{0mm}{5mm} \\ 
\end{array} \right. \label{spinor6} 
\eeq 
where $\eps_{\pm}=\frac{1}{2}(1\pm\ga)\eps$ are the chiral projections of  
$\eps$. As we will show in more detail below (see eq (\ref{log}) and the 
second equation in (\ref{eq.selfdual})), this last set of equations can  
indeed be solved on a K\"ahler manifold 
if the gauge connection cancels the $U(1)$ part of the holonomy of the 
spin connection.
We hasten to add that the physical metric is \Ka\ with 
respect to the $J_a^{\ b}$ that comes as a reduction of the octonionic 
structure constants; in general it wouldn't be \Ka\ with respect to the 
original complex structure of the Calabi-Yau.

%Notice that the normalization of the spinors $\eps_{\pm}$ does not determine
%completely the supersymmetry 
%generator $\eps$: we are left with an arbitrary constant constant phase, 
%$\eps \to e ^{iq}\eps_+ + e ^{-iq} \eps_-$. In the following we will 
%always set this phase to unity. 

In summary, we have shown that the constraints  
(\ref{6dgravitino} - \ref{6ddilatino}) for ${\cal N}=1$ supersymmetric  
compactifications of type IIA string theory to four dimensions in the 
presence of a nontrivial dilaton and RR two-form field strength
subject to the constraint (\ref{f11}), are 
realized by a warped string frame compactification (\ref{ds10})  
on a non-compact K\"ahler manifold, provided that the dilaton and  
RR two-form field strength satisfy generalizations of monopole equations 
given in (\ref{monopole.2}). The supersymmetry generator 
$\eps$ becomes a gauge covariantly constant spinor on the K\"ahler 
manifold. 
The complex structure 
$J_a^{~b}$ and the  three-form $\psi_3$ that appear in the construction 
are built using octonions. The total space of the  
''Kaluza-Klein bundle'' has a metric (\ref{ds7}) of holonomy $G_2$. 
Clearly the solution reduces to the ordinary direct Calabi-Yau  
compactification with constant dilaton if the RR two-form field strength 
is trivial, in which case it actually preserves ${\cal N}=2$ 
supersymmetry.  
 
Let us also add some comments about equation (\ref{monopole.2}).  
In the holomorphic/antiholomor\-phic basis they read
\beq 
-i \beta (\de_c \phi) = \frac{1}{8} \check{F}^{ab}\eps_{abc} 
\qquad \Leftrightarrow \qquad  
\de\left(e^{-2\beta\phi}\right)=-*\left(e^{-\al\phi}\psi_3^{(3,0)} 
\wedge F^{(2,0)}\right). 
\label{hm} 
\eeq 
Equation (\ref{hm}) closely resembles the  sort of ``holomorphic 
monopole equations'' described in \cite{Thomas,DT}, where holomorphic  
analogues of gauge theory have been studied. The definition of ``holomorphic 
monopole'' is inspired  by  the expression (\ref{hm}): when written 
in components, it is similar to the usual  
monopole, but with the 3d volume $\eps_{ijk}$ replaced by  $\eps_{abc}$ over 
holomorphic indices only. Although in our case $\eps_{abc}e^a e^b e^c$ (where 
the indices are holomorphic) is not the holomorphic three form $\Omega$ of a  
Calabi-Yau, but the $(3,0)$ part $\psi_3^{(3,0)}$ of our three-form $\psi_3$,  
we can still use it to define  a ``holomorphic  
Hodge dual'' as in \cite{Thomas,DT} 
\[ 
\star:\ \ \Omega^{(p,0)}\to \Omega^{(3-p,0)}\ ,  
\qquad \star(\alpha)\equiv *(\bar\alpha\wedge\psi^{(3,0)}) . 
\]  
Using this we can rewrite (\ref{hm}) as 
\beq 
\de\left(e^{-2\beta\phi}\right)=-e^{-\al\phi} \star F. 
\eeq 
 
It is interesting to compare (\ref{monopole.2}), (\ref{f11})  
to  the equations for gauge fields on 
D-branes wrapping CY manifolds \cite{Harvey:1996gc, Marino:2000af}. There, 
the relevant equations are known as Hermitian Yang-Mills equations, for 
which  $F^{(2,0)}$ is vanishing and $F \cdot J = {\rm const}$.  
Note that here we 
have in a way an orthogonal projection, where  $F^{(1,1)}$ is 
vanishing while equation (\ref{monopole.2})  
contains $F^{(2,0)}$ and $F^{(0,2)}$ instead.  
We will see that something of this sort also happens in the $Spin(7)\to G_2$  
case. 
 
In the next two subsections we will present alternative derivations of the
monopole equation, 
using the existence of a $G_2$ structure  
and of a selfdual spin connection  
associated with (\ref{ds7}).

\subsection{Forms and monopoles: a check} 
From the covariantly constant spinors and the gamma matrices one can, as
 usual,  form bilinear 
combinations and forms  
\beq 
J = \frac 12 J_{ab} e^a e^b  
= -\frac{i}{2}(\eps^\dagger \ga_{ab} \ga \eps)\, e^a e^b,  
\qquad  
\psi_3 = \frac{1}{3!} \psi_{abc} e^a e^b e^c  
= -\frac{i}{3!} (\eps^\dagger \ga_{abc} \eps)\, e^a e^b e^c .  
\label{psi} 
\eeq 
 
Due to the supersymmetry constraints  
(\ref{6dgravitino} - \ref{6ddilatino}), these forms satisfy a set of 
equations,  
which can be found in two ways.   
One can simply hit them with a covariant derivative, and then use 
the covariant derivative for $\eps$. Alternatively, one can  
use as a starting point the form  
\[ 
\Phi=\frac{1}{3!} \Phi_{ABC} \he ^A \he ^B \he ^C = 
 e ^{-3\al\phi} \psi_3 + e ^{-2\al\phi} J \wedge \he ^z ,
\] 
which defines the $G_2$-structure associated with the $G_2$ metric 
(\ref{ds7}). More precisely, one can reduce to six dimensions the
equations 
$d\Phi=0=d\,(*_7\Phi)$, and divide the result into pieces containing or not 
$\he^z$. Whatever way, one gets \cite{Cvetic:2002da} 
\beq 
\begin{array}{l} 
\vspace{.2cm}  
d(e^{-3\al\phi}\psi_3) + e ^{(\beta-2\al)\phi}\, J\wedge F=0\ ,  
\qquad d( e^{(\beta-2\al)\phi} J)=0\ ,\\ 
d(e^{-4\al\phi}(*J)) - e^{(\beta-3\al)\phi}\, (*\psi_3) \wedge F=0\ ,  
\qquad d( e^{(\beta-3\al)\phi} (*\psi_3))=0\ ,\\ 
\end{array} 
\label{CGLP} 
\eeq 
where the $*$ here denotes Hodge duality w.r.t.~the physical metric 
$ds_6^2$. 
 
Using the relation $\beta=2\al$ of the ten-dimensional string frame  
these equations simplify considerably. First of all the second 
equation implies that $dJ=0$. We stress once again that 
this $J$ is $\frac{1}{2}J_{ab} e^a e^b$,  
the K\"ahler form associated to the string frame 
metric $ds_6^2$ and complex structure  
$J_a^{~b}=-i\eps^{\dagger}\ga _a^{~b}\ga \eps$. There is no contradiction here 
with the fact that $J$ might happen to coincide with the \Ka\ form of the 
original Calabi-Yau metric and complex structure. 
 
To analyze these equations further and recover the monopole equations 
(\ref{monopole.2}), (\ref{f11}), we split them into bi-degrees 
w.r.t.~the complex structure $J_a^{~b}$.  
Using either the basis of spinors in 
six dimensions (see e.g.\ \cite{Becker:1995kb})  
or the octonionic structure constants in an explicit basis,  
the two-form and three-form can be written as 
\beq 
\begin{array}{r@{~=~}l} 
J&\frac{i}{2} dz^a d\bz^{\baa}, \\ 
\psi_3 & \frac{1}{2} \left( \frac{1}{3!} \eps_{abc}dz^{a}dz^bdz^{c} + 
 \frac{1}{3!} \eps_{\baa\bbb\bc}d\bz^{\baa}d\bz^{\bbb}d\bz^{\bc} \right), 
\rule{0mm}{4mm} \\ 
*\psi_3 & \frac{i}{2} \left( \frac{1}{3!} \eps_{abc}dz^{a}dz^b dz^{c} - 
 \frac{1}{3!} \eps_{\baa\bbb\bc}d\bz^{\baa}d\bz^{\bbb}d\bz^{\bc} 
\right),  
\rule{0mm}{4mm} \\ 
\end{array} 
\label{forms.hol}   
\eeq 
where $dz^a$ and $d\bz^{\baa}$ are frames on the holomorphic and 
anti-holomorphic cotangent bundle respectively.  
Notice that here we have suppressed the local phases $e ^{\pm iq}$ of
the spinors $\eps_{\pm}$ for notational convenience. 
Including them amounts to the replacements
$\eps_{abc}\to e ^{2iq} \eps_{abc}$ and 
$\eps_{\baa\bbb\bc}\to e ^{-2iq} \eps_{\baa\bbb\bc}$ respectively.

The $(3,1)$ and $(1,3)$ parts of 
the first equation in (\ref{CGLP})  
can be reduced (using also the fourth equation) 
to (\ref{monopole.2}). The third equation gives instead 
\[ 
d(*J) -4\al d\phi \wedge * J - \check{F}\wedge (* \psi_3)=0 . 
\] 
The first term vanishes since $*J=\frac{1}{2}J\wedge J$, 
and the sum of the last to 
two terms is yet another form of the holomorphic 
monopole equations. 
For integrable complex structure the $(2,2)$-part of 
the first equation in (\ref{CGLP}) reduces to  
$F^{(1,1)}\wedge J =0$, which in six dimensions is equivalent to  
$F^{(1,1)}=0$. 
Note that in the general case however, when the complex
structure is not integrable, $d\psi$ and $d(*\psi)$ also have a $(2,2)$ 
part. Both terms in the $(2,2)$ part of the first equation in
(\ref{CGLP}) are then nonvanishing. In that case $J$ is no longer a 
K\"ahler form, but still a symplectic form.

To get equation (\ref{spinor6}) back from (\ref{CGLP}) is less direct, 
the forms being bilinear in the spinor.  
Here we content ourselves with reformulating the fourth equation in 
(\ref{CGLP}) in a more suggestive way. 
Using a holomorphic basis for the vielbein 
and the hermiticity of the metric, we can define 
the quantity $e_h=\frac16\eps_{abc}\eps^{ijk}e^a_i e^b_j e^c_k$, a sort of  
holomorphic part of the determinant of the metric, such that  
$\sqrt{|g|}=e_h \bar e_h$. Then splitting the  
fourth equation in (\ref{CGLP}) gives 
\beq 
d \left(e ^{-\al\phi}(*\psi_3) \right)=0  
\quad \Leftrightarrow \quad  
\partial\left(e ^{-\al\phi}\psi_3^{(0,3)} \right) =0= 
\overline{\partial} \left(e ^{-\al\phi}\psi_3^{(3,0)} \right) 
\quad \Leftrightarrow \quad 
d \left(e ^{-\al\phi}\psi_3 \right)=0 . \label{dpsi.hol} 
\eeq 
In particular this implies 
\beq 
\bar\de\left(e^{-\al\phi}e_h\right)=0\ . 
\label{log} 
\eeq 
 
We can actually use the equations (\ref{CGLP}) to characterize the  
sub-manifold $M$ which the $N$ supersymmetric 
D6-branes wrap. Being a magnetic source of 
charge $N$ for $F$, one has 
$dF = N \delta_M$, 
where $\delta_M$ is the Poincar\'e dual three-form of the cycle $M$.   
Taking the exterior derivative of the first and third equation of  
(\ref{CGLP}) and using (\ref{dpsi.hol}) one finds that 
\beq 
J\wedge \delta_M =0, \qquad 
e ^{-\al\phi} (*\psi_3) \wedge \delta_M =0. \label{cycle.delta} 
\eeq 
Moreover, the monopole equation (\ref{monopole.2}) implies 
\beq 
\Delta \left(e ^{-2\beta\phi}\right) = 
-N *\left(e ^{-\al\phi} \psi_3 \wedge \delta_M \right). \label{lap.hol} 
\eeq  
The first equation in (\ref{cycle.delta}) means that $M$ is a 
Lagrangian cycle in the internal K\"ahler manifold. The second 
equation in (\ref{cycle.delta}) together with (\ref{lap.hol}) are what 
the additional condition of being a {\em special} Lagrangian cycle 
w.r.t.\ the Calabi-Yau structure 
turns into after the back reaction of the branes on the physical metric 
is taken into account.

\subsection{Monopoles and selfdual spin connections} 
\label{sec.monself} 
 
As in the case of the three-dimensional monopole equation (\ref{mono}), 
the monopole equations (\ref{monopole.1} - \ref{monopole.2}) can be 
traced back to the selfduality of the spin connection  
in the lift to one dimension higher.  
 
As shown in \cite{Bilal:2001an}, the $SO(7)$-gauge freedom of the 
spin connection $\haom_{AC}=\hGa_{ABC}\he ^B$  
associated with the $G_2$ metric $ds_7^2$ can be used to 
make the latter selfdual, in the sense that 
\beq 
\haom_{AB} = \frac{1}{2} (*\Phi)_{ABCD} \haom^{CD} \qquad \Leftrightarrow 
\qquad \Phi_{ABC} \haom^{AB} = 0 \ . \label{selfdual} 
\eeq 
Furthermore, this is the gauge choice 
%(with a 
%particular representation of $\ga$-matrices) 
for which  
the covariantly constant spinor $\eps$ is actually  
constant and has a single nonvanishing entry, namely 
the singlet in  
$\underline{8} \rightarrow \underline{7} + \underline{1}$. 
 
Equation (\ref{selfdual}) implies 
\[ 
0=\Phi_{ab7}\haom^{ab}= J_{ab} \left( 
 \om^{ab} +2\al(\de ^a \phi) e ^b -\frac{1}{2} e ^{\al\phi} 
 \check{F}^{ab} \he ^z \right) 
\] 
and 
\bean 
0&=&\Phi_{ABc}\haom^{AB}=\psi_{abc}\haom^{ab}+2J_{bc} \haom^{7b} \\  
&=& \psi_{abc} \left( 
 \om^{ab} + 2\al(\de ^a \phi) e ^b -\frac{1}{2} e ^{\al\phi} 
 \check{F}^{ab} \he ^z \right) 
 +2J_{bc} \left( \frac{1}{2}  
 \check{F}^{b}_{~d} e ^d + \beta e ^{\al\phi} (\de ^b \phi) \he ^z 
\right).  
\eean 
Decomposing these equations into the basis $e ^a,\he ^7$ we find the  
following constraints 
\bea 
0&=& J_{ab} F^{ab} , \non \\ 
0&=& J_{ab} \left( \G^{a~b}_{~c} + 2\al\delta_c^{~b}(\de ^a \phi) 
 \right) , \non \\ 
0&=& \beta (\de ^b \phi) J_{bc} - \frac{1}{4}  
 \check{F}^{ab} \psi_{abc} , \label{eq.selfdual} \\ 
0&=& \check{F}_d^{~b} J_{bc} - \left( 
 \G^{a~b}_{~d} + 2\al\delta_d^{~b}(\de ^a \phi) \right) \psi_{abc} . \non 
\eea 
  
The same equations also follow from the type IIA supersymmetry 
constraints (\ref{6dgravitino} - \ref{6ddilatino}). We recall that in 
the $SO(7)$-gauge in which the spin connection $\haom_{AB}$ is 
selfdual, the spinor $\eps$ has a single constant component. 
In this gauge the derivative term in  
(\ref{6dgravitino}) drops out, and, using
(\ref{6dga}), we can rewrite (\ref{6dgravitino} - \ref{6ddilatino}) as
\beq 
0= \left\{ \left[ \left( 
 \G^{a~b}_{~d}+2\al\delta_d^{~b}(\de ^a\phi) \right) \psi_{abc} -  
 \check{F}_d^{~b}J_{bc} \right] \ga ^c  
 + \left[ \left(\G^{a~b}_{~d}+2\al\delta_d^{~b} 
 (\de ^a\phi) \right) J_{ab} \right] 
 \ga \right\} \eps \label{var.grav.frame.6c} 
\eeq 
and  
\beq 
0= \left\{ \left[ - \beta (\de ^b \phi) J_{bc} 
 + \frac{1}{4} \check{F}^{ab} \psi_{abc} \right] \ga ^c 
 + \left[ \frac{1}{4} \check{F}^{ab} J_{ab} \right] 
 \ga \right\} \eps \ .  \label{var.dil.6c} 
\eeq 
Decomposing into a basis 
$\eps, \ga ^A \eps$ for the spinors, we obtain the same constraints 
(\ref{eq.selfdual}) as by using the selfduality of the spin connection. 
 
The first and third equation in (\ref{eq.selfdual}) give once again 
the monopole equations (\ref{monopole.1} - 
\ref{monopole.2}). Inserting the third equation into the fourth and 
using the string frame relation $\beta=2\al$ one derives the 
constraint 
\beq
\G^{a~b}_{~d}\psi_{abc}= \frac{1}{2} 
 \left(\check{F}_{cb}J^b_{~d} + \check{F}_{db}J^b_{~c} \right)
 . \label{Gpsi} 
\eeq
Under our assumption $F^{(1,1)}=0$, (\ref{f11}), the right hand side
is identically zero. $\G^{a~b}_{~d}\psi_{abc}=0$ then implies the projection 
of the general $SO(6)$ holonomy of the spin connection 
$\underline{15}=\underline{9}+\underline{3}+\underline{\bar{3}} 
\mapsto \underline{9}$ into $U(3)$ holonomy. Hence the 
geometry is K\"ahler. 

Finally, the second equation in (\ref{eq.selfdual}) 
implies that $2\al(\de ^a \phi)J_{ac}$ defines a $U(1)$ gauge 
connection for the six-dimensional spinor that cancels the $U(1)$ part 
of the holonomy of the spin connection. Indeed, the term 
$J_{ab}\G^{a~b}_{~d}$  
is  
the $U(1)$ part of the connection. Due to K\"ahlerity,  
the connection has a particularly compact expression, that allows to  
rewrite the second equation as  
\[ 
E^{\bar a k}\bar\de_{\bar j} e^a_{\ k}= \al \bar\de_{\bar j} \phi\ , 
\] 
where $E$ is the inverse vielbein. Now, the first piece is nothing but 
$Tr{E^{-1}\bar\de_{\bar j} e}=\bar\de_{\bar j}(\log e_h)$; so we get  
$\bar\de_{\bar j}(\log e_h -\al \phi)=0$,  
which is (\ref{log}). Finally, we are now able to show more directly the  
connection between (\ref{spinor6}) and the  K\"ahlerity condition.
As we have just done for 
(\ref{var.grav.frame.6c}),  
use the 
action of the gamma matrices on the spinor and gauge  
the latter to be a  
constant: it is easy to see that K\"ahlerity follows, in the form  
$\G^{a~b}_{~d}\psi_{abc}=0$, and one is left with the second equation in  
(\ref{eq.selfdual}).

\newsection{$Spin(7) \to G_2$} 
\label{sec.spin7g2}

This case will not be treated in as much detail as the previous one; 
we will only show that a ``$G_2$ monopole equation'' arises. 
The equation in seven dimensions that one gets from the 
reduction of eight dimensions is formally the same as in  
(\ref{6dgravitino}-\ref{6ddilatino}), but for the fact that $\gamma$ 
is now the identity and can be dropped, $\epsilon$ being now a spinor in 
seven dimensions which was a Weyl spinor in eight dimensions.  
%\beq 
%\left( D_A + \frac\al2 \de_B \phi \ga^B_{\ A}+ \frac i4 \check{F}_{AB} \ga 
%  ^B\right) \eps =0 \ . 
%\eeq 
A basis for spinors is then 
spanned by  $\eps$ and $\ga ^A \eps$ \cite{Marino:2000af}; 
moreover we have  
\beq 
\ga_{AB}\eps= i\Phi_{ABC} \ga ^C \eps\ . 
\eeq 
Using this, the gravitino and dilatino equations become 
\bea 
&&D_A \eps + i\left[\frac\al2(\de ^B \phi) \,\Phi_{BAC} 
  + \frac 14 \check{F}_{AC} \right] \ga ^C\eps =0\ .\\ 
&&i\left[ \frac14 \check{F}^{AB} \,\Phi_{ABC} + \,\beta \,\de_C\phi \right]  
  \ga ^C \eps=0.\label{g2.dil} 
\eea 
 
The dilatino equation gives us again a monopole equation. 
Reinserting this equation back 
into the gravitino one, the latter becomes
\beq
D_A \eps = -\frac{i}{4} \left[ \left(1-\frac{\al}{\beta}\right) 
 \check{F}_{AC} + \frac{\al}{2\beta} (*\Phi)_{ACDE}\, \check{F}^{DE} 
 \right] \ga ^C \eps . \label{g2.grav.b}
\eeq
In a frame with $\beta=3\al$ the right hand side vanishes if
\beq 
\check F_{AB} +\frac 14 (*\Phi)_{ABCD} \,\check F^{CD}=0 . 
\label{g2.proj} 
\eeq 
The last condition 
means that in the reduction $\underline{21} \rightarrow 
  \underline{14} + \underline{7}$ of the adjoint of $SO(7)$
under its subgroup $G_2$, we  
project out the fields in the ${\underline{14}}$,  
whereas for $G_2$ instantons it is the ${\underline{7}}$ that is 
projected out \cite{Acharya:1997jn}. 
This condition is the analogue of 
the projection (\ref{f11})
in six dimensions, where $F$ was assumed 
to have components only in the representations 
$\underline{3}+\underline{\bar 3}$, orthogonal to what one has for 
Hermitian Yang-Mills equations (see the remark after (\ref{hm})). 

Assuming (\ref{g2.proj}) to hold, the gravitino equation requires the 
physical metric in the frame $\beta=3\al$ to allow for a covariantly 
constant spinor and hence to be of holonomy $G_2$. This in turn implies
that the physical string frame metric, for which $\beta=2\al$, is 
conformal to a $G_2$ metric,  
\beq 
ds_7^2 = e ^{\frac{2}{3}\al\phi} ds_{G_2}^2 . 
\label{conf.g2} 
\eeq 

Analogously to the subsection \ref{sec.monself},
the same conclusions can again be drawn from the existence of a selfdual
spin connection on the eight-dimensional lift. In that gauge the spinor
is constant. Whereas the dilatino equation (\ref{g2.dil}) stays unchanged,
the gravitino equation, after reinserting the dilatino one, becomes 
\beq
\G^{B~D}_{~A}\Phi_{BDC} = -\left[ \left(1-\frac{\al}{\beta}\right) 
 \check{F}_{AC} + \frac{\al}{2\beta} (*\Phi)_{ACDE}\, \check{F}^{DE} 
 \right] . \label{g2.grav.c}
\eeq 
Assuming (\ref{g2.proj}) one sees that the spin connection coefficients in
the frame $\beta=3\al$ satisfy the selfduality condition 
$\G^{B~D}_{~A}\Phi_{BDC} =0$, so that the physical metric in this frame has
$G_2$ holonomy.
 
It is also not difficult to derive the analogue of (\ref{CGLP}) from 
the fact that  
%the eight dimensional $\Phi_8$ is closed;  
the lift to eight dimensions has a $Spin(7)$-structure,
\beq
\Omega_8 = - e ^{-4\al\phi}(*\Phi)-e ^{-3\al\phi} \Phi \wedge \he ^z .
\eeq
Reducing $d\Omega_8=0$ to seven dimensions  
one gets 
\beq 
d(e^{-4\al\phi}\,(*\Phi))= 
e^{(-3\al+\beta)\phi}\,\Phi\wedge F\ ,\qquad 
d(e^{(-3\al+\beta)\phi}\Phi)=0\ . 
\label{CGLPG2} 
\eeq 
Again, if we suppose $3\al=\beta$, the 
second equation yields $d\Phi=0$. The first one can be rewritten as  
\[ 
d(*\Phi)-4\al d\phi\wedge*\Phi- \check F\wedge \Phi=0\ ,  
\] 
and the most natural way to solve it is to set the sum of the last two  
terms to zero, which gives our monopole equation coming from the dilatino, 
(\ref{g2.dil}), and $d(*\Phi)=0$. The latter together with $d\Phi=0$ 
implies once again that the metric in the frame $\beta=3\al$ has a 
holonomy contained in $G_2$. 
 
From (\ref{CGLPG2}) we can again characterize the submanifold $M$ 
on which $F$ has a source, $dF=N\delta_M$. Hitting the first equation 
with $d$ and using the second, we obtain the condition  
\[ 
e^{(-3\al+\beta)\phi}\Phi\wedge\delta_M=0\ , 
\] 
which means that the cycle is coassociative with respect to the 
$G_2$ structure $\Phi$ in the frame $\beta=3\al$. Likewise,  
the monopole equation 
(\ref{g2.dil}) can be rewritten as $d(e^{-4\alpha \phi})=\frac23 
*(F\wedge *\Phi)$, or  
\beq 
\Delta\,( e^{-4\al \phi})= \frac23 N *(\delta_M\wedge*\Phi)\ . 
\eeq 
 
\newsection{Conclusion} 
\label{sec.disc}

We have seen how the conditions for preserving supersymmetry in type IIA 
string theory in the presence of a nontrivial dilaton and RR two-form 
field strength get reduced to generalizations of the monopole equations. 
The essential ingredient of this generalization is the existence of a 
three-form in the underlying geometry. 
The special role played by 
three-forms in describing six- and seven-dimensional structures is 
well-studied \cite{Hitchin:2000jd}. Here our emphasis has been on  
the circle fibrations on such spaces, and we  
have shown that in order for the total space of such a ``Kaluza-Klein circle 
bundle'' to yield a manifold of restricted 
holonomy, the volume of the circle, the field strength of the 
Kaluza-Klein vector and the three-form have to satisfy generalized 
monopole equations.  
If moreover the field strength has additional projection properties, 
(\ref{f11}) or (\ref{g2.proj}), the base manifolds with their string frame
metric are K\"ahler for $d=6$ and conformal to a $G_2$ manifold for
$d=7$. 

It would be desirable to characterize the base geometries further, even
if these supplementary constraints are not imposed.
This even more, since they prohibit 
an easy inclusion of the cases with lower $d$
into those for higher $d$, as can be done for vanishing background fluxes, 
where we have successive inclusions of the holonomies,
$\{1\}\subset SU(3) \subset G_2$ and 
$SU(2)\subset G_2 \subset Spin(7)$, respectively. 
For example the known solution for D6-branes in flat space, described in
section \ref{sec.flat}, satisfies the conditions (\ref{6dgravitino} - 
\ref{6ddilatino}) for ${\cal N}=1$ supersymmetry in four dimensions,
or alternatively their form 
(\ref{monopole.1} - \ref{6dgrav.b}). However, w.r.t.\ the almost
complex structure that descends from the octonions, $F^{(1,1)}$ is
nonvanishing. This almost complex structure is therefore not integrable
and the string frame metric on the internal six-dimensional space not
K\"ahler. 

A better understanding of the general base geometries and the conditions  
on the connections on the  circle bundles should hopefully lead to some  
systematic construction of new metrics of restricted holonomy.  
One may apply a warping
$ds_6^2 = e ^{\beta\phi} ds_{3} + 
e ^{-\beta\phi} d\tss_{3}$, 
analogous to the flat case, to a 
cotangent bundle of any three-dimensional manifold. The latter are local 
realization of SLAGs inside a Calabi-Yau threefold. If one is able to solve
(\ref{monopole.1} - \ref{6dgrav.b}) for $\phi$ and $A$, one could get 
at least local expressions for new metrics of $G_2$ holonomy.

We would like to conclude this paper by mentioning another direction for 
further research that we have not pursued here. The possibility of having 
more general setups where other RR fluxes are turned on is of obvious 
physical interest. In some cases this lifts to internal  
manifolds of weak restricted holonomy, on which 
spinors instead of being covariantly constant satisfy $D_A \eps = 
i\,\lambda\,\ga_A \eps$. Further deformations of both the underlying geometry 
and the monopole equations arising in this context have been left for future 
work.

\vspace{.2cm}  
\section*{Acknowledgments} 
 
We would like to thank B. Acharya, D.~Calderbank, M.~Douglas, 
D. Martelli for useful discussions. 
This work is supported in part by EU contract HPRN-CT-2000-00122 and by 
INTAS contracts 55-1-590 and 00-0334. PK and MP are supported by 
European Commission Marie 
Curie Postdoctoral Fellowships under contract numbers  
HPMF-CT-2000-00919 and HPMF-CT-2001-01277. 
%\vspace{2cm} 
 
\pagebreak

\end{document}